\documentclass[twocolumn]{article}

\usepackage[pdftex]{graphicx}
\usepackage{amsmath,amssymb}
\usepackage{hyperref}

\title{Sonovortex: Aerial Haptic Layer Rendering by~Aerodynamic Vortex and Focused Ultrasound}

\author{Satoshi Hashizume\thanks{hashizume@digitalnature.slis.tsukuba.ac.jp} \and Amy Koike \and Takayuki
Hoshi \and Yoichi Ochiai \\ \and University of Tsukuba}

\date{}

\begin{document}
\maketitle

\begin{abstract}
In this paper, a method of rendering aerial haptics that uses
  an aerodynamic vortex and focused ultrasound is presented.
  Significant research has been conducted on haptic applications based on multiple phenomena such as magnetic and  electric
 fields, focused ultrasound, and laser plasma. 
By combining  multiple physical quantities; the
 resolution, distance, and magnitude of force are enhanced. To combine multiple tactile technologies,
 basic experiments on resolution and discrimination threshold are required.
Separate user studies were conducted using 
  aerodynamic and ultrasonic haptics. 
Moreover, the perception of their superposition, in addition to their resolution, was tested.
  Although these fields cause no direct
  interference, the system enables the simultaneous perception of the tactile feedback of
  both stimuli. The results
  of this study are expected to contribute to expanding the expression of aerial haptic
  displays based on several principles.

\end{abstract}

\section{Introduction}
Aerial haptic feedback is a popular topic in research fields such as 
real-world-oriented interaction, augmented reality (AR), and virtual
reality (VR). Several methods have therefore been proposed to realize aerial haptic feedback
which include phenomena such as magnetic forces, ultrasound,
and air vortices.

An aerial haptic display has several advantages. First, it projects a
force over a distance without physical contact or wearable
devices. Second, it has a high programmability. In particular, it
can be set and rearranged at an arbitrary position in a three-dimensional (3D)
space, as it does not require physical actuators.

\begin{figure}[tb]
\centering
 %\rule{7truecm}{5truecm}
 \includegraphics[width=\linewidth]{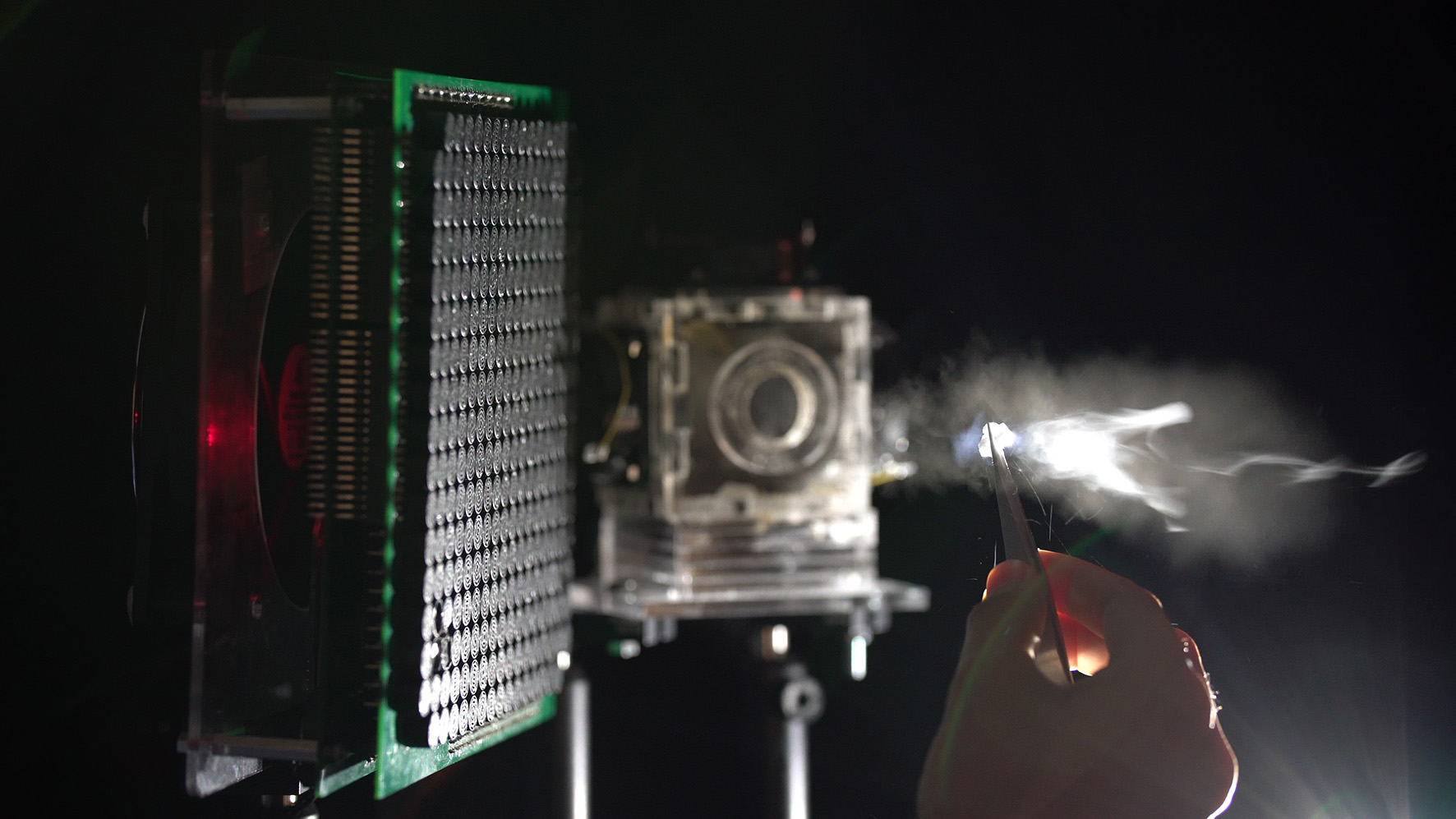}
  \caption{Sonovortex device.}
 \label{fig:teaser}
  \end{figure}

In this study, new aerial interactions were evaluated.
 The aim of the study was the development of  a new aerial haptics system to express
a wide range of feedback.
The proposed system~(Figure \ref{fig:teaser}) combines aerodynamic and acoustic fields.
 The aerodynamic vortex~\cite{Weigand1997} from
the aerodynamic field and the focused ultrasound~\cite{5406524} from the acoustic field were used to develop the device.

The tactile sensations of  single and multiple
fields were then compared. Through a user study, it was found that the aerodynamic
vortex and focused ultrasound do not influence each other. 
 By combining different types of forces, the proposed system can display various textures.
Based on the reports in the literature, this is an early study that combines
multi-field physical quantities to render haptic textures.

\section{Related Work}
 \subsection{Aerial Haptics Feedback}
Several methods have been proposed for aerial haptic feedback without
physical contact or wearable devices.
The technologies employed without wearable devices are based on
aerodynamic vortices, focused ultrasound, laser induced plasma, and
magnetic forces. These technologies have been applied to touch
panels~\cite{Monnai:2014:HMH:2642918.2647407} and VR~\cite{Suzuki:2005:AJD:1042190.1042215} systems . 

Ultrasonic technology, which uses ultrasound, can provide tactile sensations in mid-air
without the need for a user actuator~\cite{Carter:2013:UMM:2501988.2502018} \cite{5406524}. The position of
the focal point can be changed using a phased array transducer
as it  represents tactile
behavior. The rendering of volumetric haptic shapes can also be achieve using focused ultrasound~\cite{Long:2014:RVH:2661229.2661257}.
MidAir~\cite{Monnai:2014:HMH:2642918.2647407} reflects a
virtual image in the air and provides tactile feedback 
using an ultrasonic speaker based on the virtual image and
finger location. HaptoClone~\cite{Makino:2016:HMT:2858036.2858481}
enables real-time interaction with floating volumetric images using haptic
feedback. This method is insufficient, given that the focused ultrasound force is very weak and
only limited focal points can be generated.

In \cite{Suzuki:2002:DFF:506443.506608}, an air jet was used to produce contactless haptic
feedback with a low accuracy. In~\cite{Suzuki:2005:AJD:1042190.1042215},
virtual objects were represented by air jets from an array of nozzles.
Vortex rings~\cite{Weigand1997} have also been used as a non-contact
haptic feedback system, and  air vortices have been
used to produce impact in midair~\cite{Sodhi:2013:AIT:2461912.2462007}
\cite{Gupta:2013:ANH:2493432.2493463}. These previous
approaches mainly used vortex rings to add multi-modal sensation to a
conventional display. Ueoka et al.\cite{Ueoka:2016:ICH:2851581.2892299}
evaluated the manner in which people perceive haptic stimuli generated by air vortex
rings and how the stimuli affect their emotional states when
stressed. However, both these methods have a low
fidelity due to the non-focusing stimulating area.

The tactile presentation of a magnetic field involves both direct
and indirect presentation. For direct tactile feedback, a magnet can be placed on the
finger~\cite{Weiss:2011:FNH:2047196.2047277} .
The authors in~\cite{Karunanayaka2013}, presented magnetic based sensing in
addition to haptic feedback. Zhang et al.~\cite{7372383} and 
Berkelman et al.~\cite{6183773}
rendered a 3D model in mid-air using an electromagnet array.
In direct tactile presentation, powerful tactile feedback can be achieved
 without touching the screen.

Light is employed to provide  sensation on the hands when the user
is experiencing thermal radiation~\cite{Saga2015}. Nanosecond lasers
applied to the skin induce a tactile
sensation~\cite{Ochiai:2016:FLF:2882845.2850414} . To date, radio-frequency and superconducting forces
have not been applied to aerial haptic feedback.

 \subsection{Cross-Field Haptics}
This study combines multiple haptics technologies, thus overcoming their individual drawbacks and improving
the interaction width.

The wUbi-Pen~\cite{Kyung:2008:WWG:1401615.1401657} is a pencil-type tactile interface that
consist of a vibrator, linear vibrator, speaker, and pin array.
It provides functions such as feedback drag, drop, and
movement. Minamizawa et al.~\cite{5444646} developed a device based on tactile
presentation, which that combines one-point kinesthetic feedback
and multipoint tactile feedback. The accuracy of the
feedback was then improved by combining haptics
technologies. Impacto~\cite{Lopes:2015:ISP:2807442.2807443} was 
designed to render the haptic sensation of
hitting and being hit in the  VR environment. They combined tactile
stimulation with electrical muscle stimulation. Cross-Field Aerial
Haptics~\cite{Ochiai:2016:CAH:2858036.2858489} involves the drawing of  a tactile interface in
the air by combining ultrasonic waves and laser plasmas. Hashizume
et al.~\cite{7989930} developed a touch type
haptic device that combines magnetic and electrostatic fields. Their
report also includes a description of their implementation method. 

Cross-field haptics is not a widely studied field. 
Aerodynamic vortex and focused ultrasound are simultaneously utilized. 
Using both fields helps eliminate their drawbacks and
it is intended to provide a wider tactile presentation. The aerodynamic
vortex provides tactile sensations over large distances and forces.
The focused ultrasound delivers distinguishable high-resolution tactile
sensations.

\section{Implementation}
\begin{figure}[tb]
\centering
  %\rule{17truecm}{5truecm}
  \includegraphics[width=0.8\linewidth]{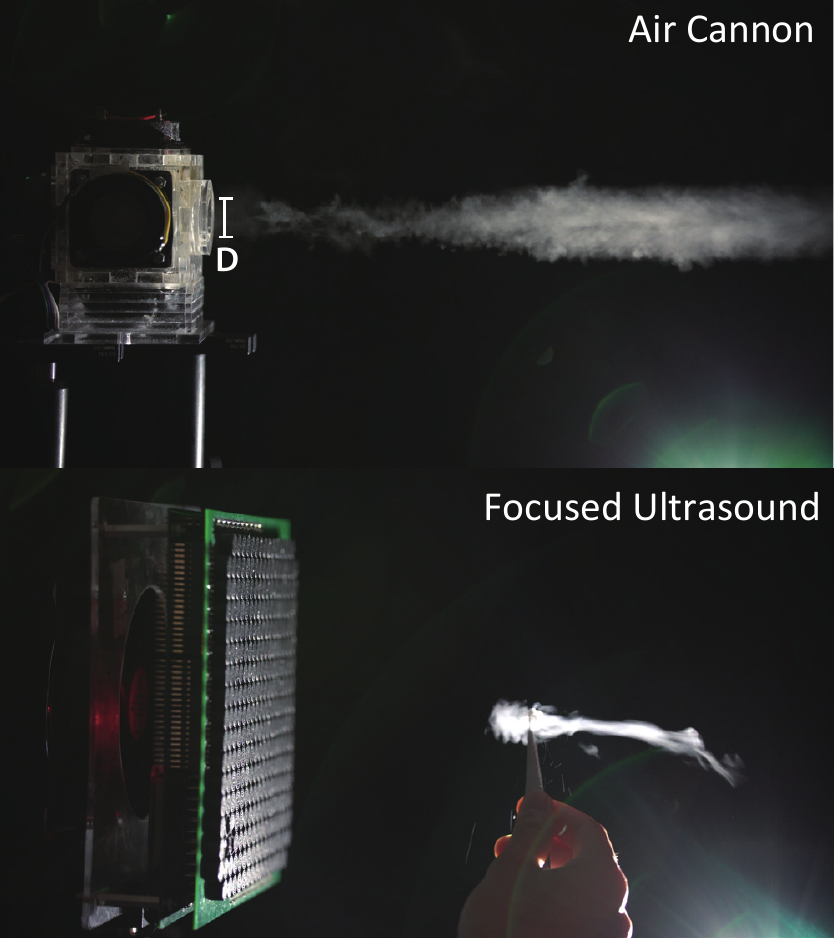}
  \caption{Air cannon and focused ultrasound visualized using dry ice and smoke.}
\label{fig:smoke}
 \end{figure}
\begin{figure*}[tb]
\centering
  %\rule{17truecm}{5truecm}
  \includegraphics[width=0.8\linewidth]{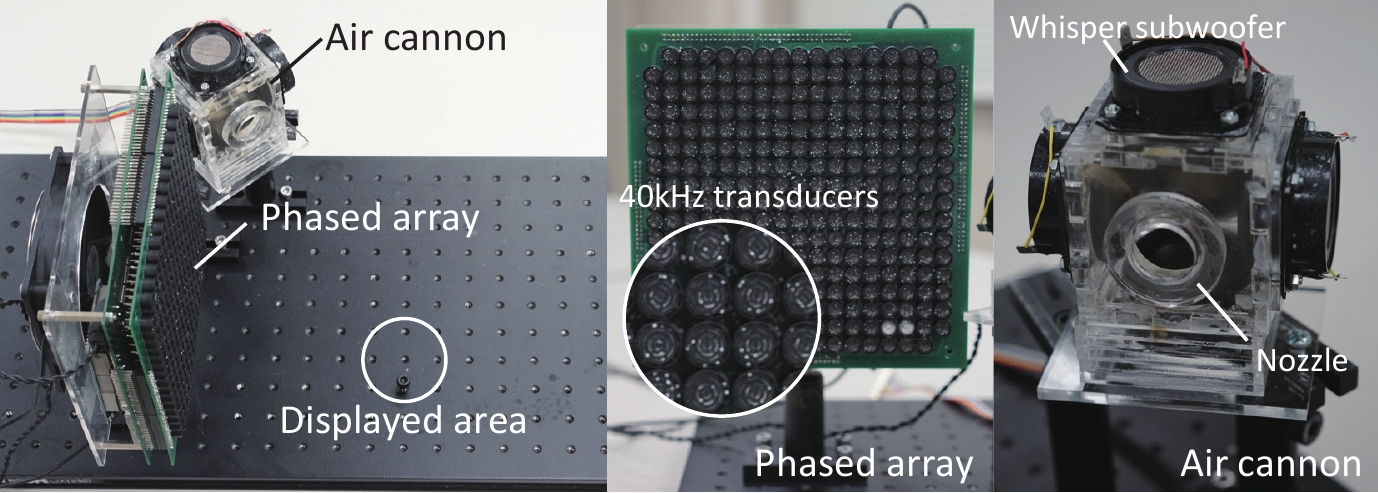}
  \caption{Aerodynamic and ultrasound system.}
  \label{fig:system}
 \end{figure*}
 \subsection{Aerodynamic Haptics}
 An air vortex (Figure \ref{fig:system}, right) is a ring of air that typically has a toroidal shape and
can travel at high speeds over large distances. 
Vortex rings (Figure \ref{fig:smoke}, upper)  can be formed by pushing air using a piston through a
circular aperture, or hole. The quality of the formed vortex is dependents on the
volume of air pushed, the velocity of the piston, and the diameter of
the aperture.

Gharib et al.~\cite{Weigand1997} defined the stroke ratio $R_{Stroke}$ 
as a ratio of the length of the theoretical cylindrical slug of air
pushed out of the nozzle $L_S$ to aperture of diameter $D$:
\begin{equation}
 R_{Stroke} = \frac{L_S}{D} \nonumber
\end{equation}
The stroke ratio characterizes the stability of the vortex as it
exits the aperture, and it is used to define the formation number.
A typical value for the formation number is within the range of 3.6-4.5 
for several various vortex systems.

According
to~\cite{doi:10.1146/annurev.fl.24.010192.001315}\cite{:/content/aip/journal/pof1/31/12/10.1063/1.866920},
$L_S$ can then be  expressed as
\begin{equation}
 L_S = \frac{4 V_S}{\pi D^2} \nonumber
\end{equation}
where $V_S$ is the slug volume.

Thus, for stable vortices, the following is true:
\begin{equation}
 \frac{4 V_S}{\pi D^3} \leqq 4.5
\end{equation}

An air cannon based on AIREAL, was developed in this study~\cite{Sodhi:2013:AIT:2461912.2462007}.
Five 2-inch 15W Whisper subwoofers were used as actuators.
Sodhi et al.~\cite{Sodhi:2013:AIT:2461912.2462007} determined 
the total volume of air displaced by all five speakers and the aperture
diameter using the previous equations:
\begin{equation}
 V_S = 33,670 \, \mathrm{mm^3}, D \geqq 2.1 \, \mathrm{cm} \nonumber
\end{equation}

  \subsection{Ultrasound Haptics}
Ultrasonic haptics are based on acoustic radiation
pressure, which exerts a force on the surface of the skin
(Figure \ref{fig:smoke}, lower). Ultrasonic haptics can be applied to the skin for a long
time-period; however,
they are relatively weak (10-20 mN). The sensation is similar
to that of a laminar air flow within a narrow area.

The time delay $\Delta t_{ij}$ for the $(i, j)$-th transducer is
given by:
\begin{equation}
 \Delta t_{ij} = \frac{l_{00} - l_{ij}}{c}
\label{eq:trans}
\end{equation}
where $l_{00}$ and $l_{ij}$ are the distances from the focal point to 
the $(0, 0)$-th reference and $(i, j)$-th transducers, respectively; and
$c$ is the speed of sound in the air. The focal point can be moved by
recalculation and the setting of the time delays for the next coordinates.

Haptic images are generated by an acoustic
phased array system(Figure \ref{fig:system}, middle). Haptic image Hi is the summation of
the time series of the focal points:
\begin{equation}
 H_i = \sum{f_p(x, y, z) \times p \times t} 
\end{equation}
where $f_p$ is the ultrasonic focal points, $p$ is the acoustic pressure, and $t$ is the
time duration.

  \subsection{Latency}
The amount of time required to produce a tactile presentation using an aerodynamic
vortex is different from that required using the focused ultrasound. The focused ultrasound advances in
the air at the speed of sound. Therefore, the focused ultrasound can achieve
a near simultaneous tactile presentation with the generation of
signals. The speed and delay of the aerodynamic vortex are greater than
those of ultrasound. According
to~\cite{Gupta:2013:ANH:2493432.2493463}, the vortex speed is given by
\begin{eqnarray}
 v_{s} = \frac{L_S}{t_{cone}} \nonumber \\
 v_{vortex} = \frac{v_s}{2} \nonumber
\end{eqnarray}
where $t_cone$ is the time required for the displaced air to move through the
aperture. The average speed of the aerodynamic vortex device used in this study was 
$7.2m/s$~\cite{Sodhi:2013:AIT:2461912.2462007}. A delay of 30mms was applied to  the focused
ultrasound to ensure that the aerodynamic vortex and focused
ultrasound reach the user simultaneously.

\section{EXPERIMENTS AND RESULTS}

In this section, a discussion on the user experiments for the evaluation proposed haptic system is presented. 

\subsection{Experiment of generated force}

\begin{figure}[tb]
\centering
  \includegraphics[width=\linewidth]{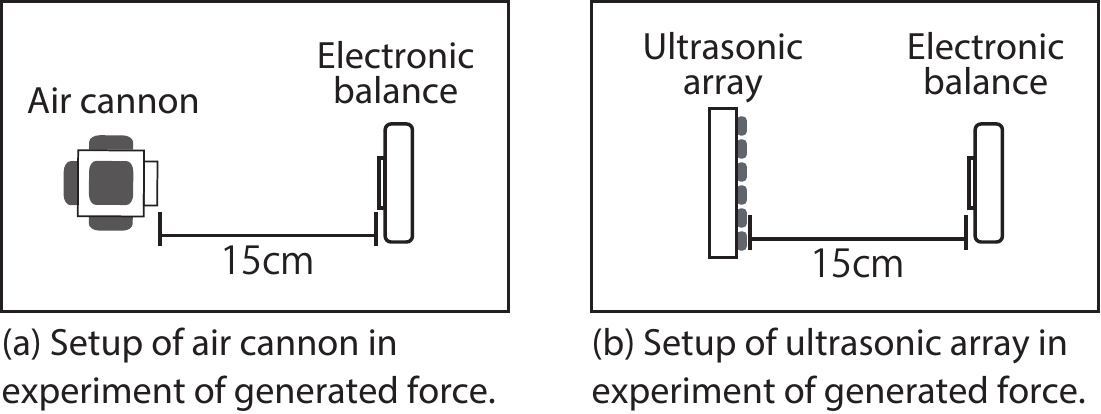}
  \caption{Setup of experiment on generated force}
\label{fig:setup_force}
 \end{figure}

The magnitude of the force generated by an ultrasonic wave and air
cannon were considered. The precision electronic balance was set  vertically and placed 15 cm from the ultrasonic
device and air cannon (Figure \ref{fig:setup_force}). The mass displayed on the precision electronic balance
was then converted to
force. The air cannon was output at 30 Hz, and the power supply voltage was varied from 5 to 17.5
V in increments of 2.5 V. In addition the change in the magnitude of the generated force was examined. The
ultrasonic wave was output at a modulation frequency of 50 Hz, and the output intensity was
changed from 0 to 600 in increments of 100. Moreover, the change in the magnitude of the generated force
was investigated. To convert the output intensity to the output force, a conversion could be carried out
using $\sin^2(\pi p / 1248)$~\cite{5406524}. The ultrasonic focal length was set as 15 cm.

\begin{figure*}[tb]
\centering
  \includegraphics[width=0.8\linewidth]{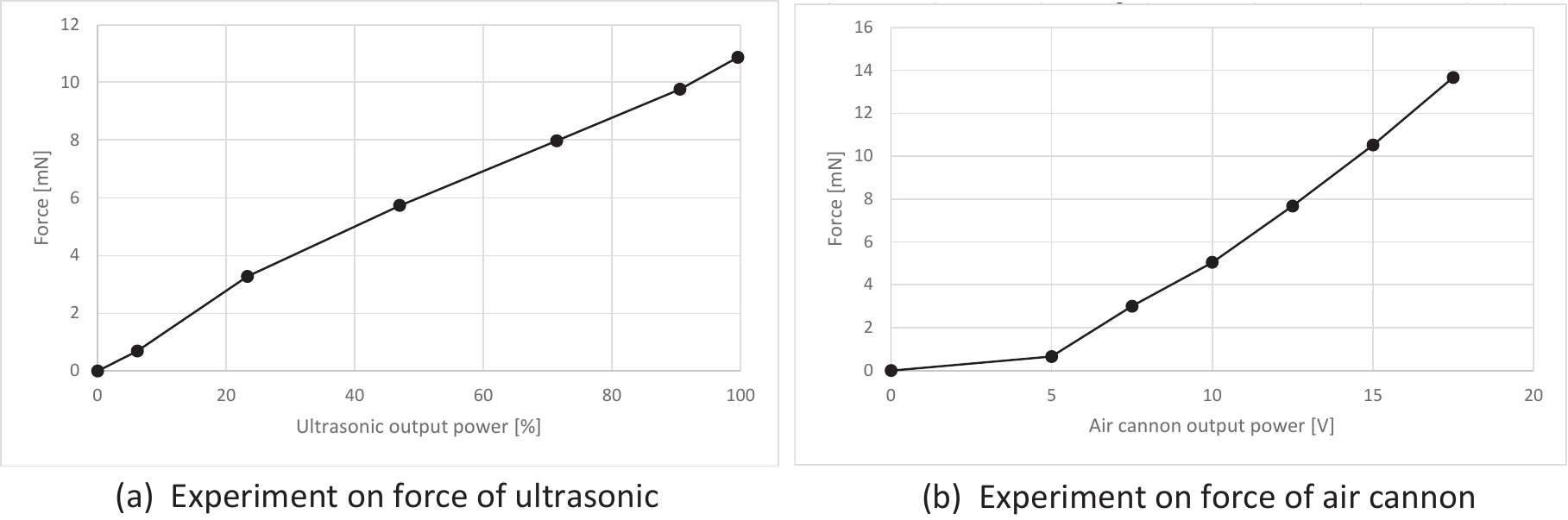}
  \caption{Results of experiment on generated force}
\label{fig:result_force}
 \end{figure*}

Figure \ref{fig:result_force} present the results, in which both the ultrasonic wave and air cannon increase the force generated in proportion to the output intensity.

 \subsection{Experiment of double-point threshold}

The user study was conducted to investigate spatial resolution.
The double-point threshold\cite{dellon1987reliability} for acoustic radiation pressure induced by focused
ultrasound and air vortex pressure was evaluated. Five people participated in
the user study (20.2 years old on average, with one female and four males).
The participants were isolated from visual information using blindfolds.
Participants then placed their hands on a table positioned 15 cm away from the haptic device(Figure \ref{fig:expsetup} right). 
The platform could be moved with an accuracy of 0.1 mm, as participants were told not to move their hands.
The output force was set with reference to Figure \ref{fig:result_force}.

The method of limits was used to measure the double-point threshold. First, the standard
stimulus was applied to the palm of the hand of the participant. 
The standard stimulus is a 10 s ultrasound wave and five air vortex cycles.
After administering the standard stimulus, the platform was moved and the comparative stimulus was administered.
After administering the comparative stimulus, each participant gave one of the following answers:
"(1) the two 
points are divided," (2) " the two points not are divided," or (3) "I do not know." The distance between the standard and
the comparative stimuli was approximated until the participant answered "the two points are not
divided" or "I do not know" (descending series). The distance between the standard 
and comparative stimuli was then increased until the participant answered with "the two points are divided"
(ascending series). The test for the descending and ascending series were carried out twice.

The experiments were conducted when only ultrasonic waves were applied, when only an air cannon was
used, when an air cannon was used with a constant  ultrasonic wave, and
when an ultrasonic wave was applied with a constant  air vortex.
The ultrasonic waves were generated at modulation frequencies of  50 Hz and 200 Hz. 
The focal length of the ultrasonic device was set as 15 cm, and the output force  was set as 5.73 mN.
The participants were stimulated with ultrasound for 5 s.
The output force of the air cannon was set as 7.67 mN, and the stimulation was applied five times. 
To provide a constant  air vortex, the air cannon was implemented at 15 Hz.
  \begin{figure}[tb]
\centering
 % \rule{17truecm}{8truecm}
\includegraphics[width=\linewidth]{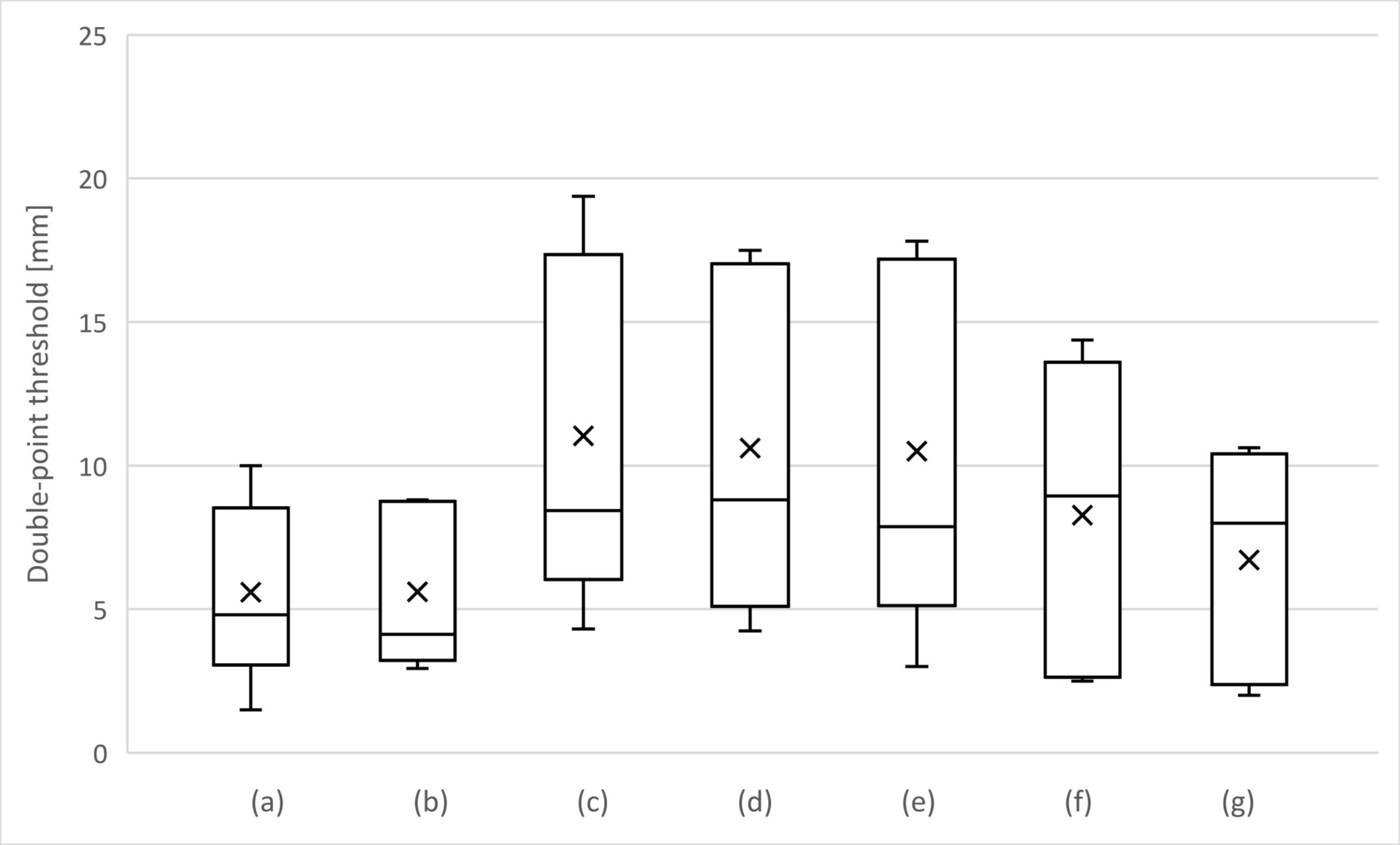}
  \caption{Results of  double-point threshold: (a)ultrasonic waves(50Hz), (b)ultrasonic waves
  (200Hz),  (c)an air cannon, (d) an air cannon while ultrasonic wave (50Hz) was constantly
provided, (e) an air cannon while ultrasonic wave (200Hz) was constantly
procided, (f) ultrasonic wave(50Hz) while an air vortex was constantly provided. (g) ultrasonic wave(200Hz) while an air vortex was constantly provided.}
\label{fig:result_pointth}
 \end{figure}

{\bf Results} : The results are shown in Figure \ref{fig:result_pointth}.
In the case of only ultrasonic waves (Figure \ref{fig:result_pointth} (a) and (b)), the double-point
threshold was approximately 6 mm regardless of the modulation frequency. The double-point threshold for an air cannon
only (Figure \ref{fig:result_pointth} (c)) was 11 mm. The double-point threshold of an air cannon
while an ultrasonic wave was constantly provided (Figure \ref{fig:result_pointth} (d) and (e)) was not much
different from that of an air cannon only. The double-point threshold of the
ultrasonic wave while an air vortex was constantly provided (Figure \ref{fig:result_pointth}
(f) and (g)) was approximately 3 mm larger that of the ultrasonic only. In the case in which the
air cannon was affected, the variation of the double-point threshold by the participant was large.

 \subsection{Experiment of perceptual threshold}
 \begin{figure}[tb]
\centering
  %\rule{17truecm}{5truecm}
  \includegraphics[width=\linewidth]{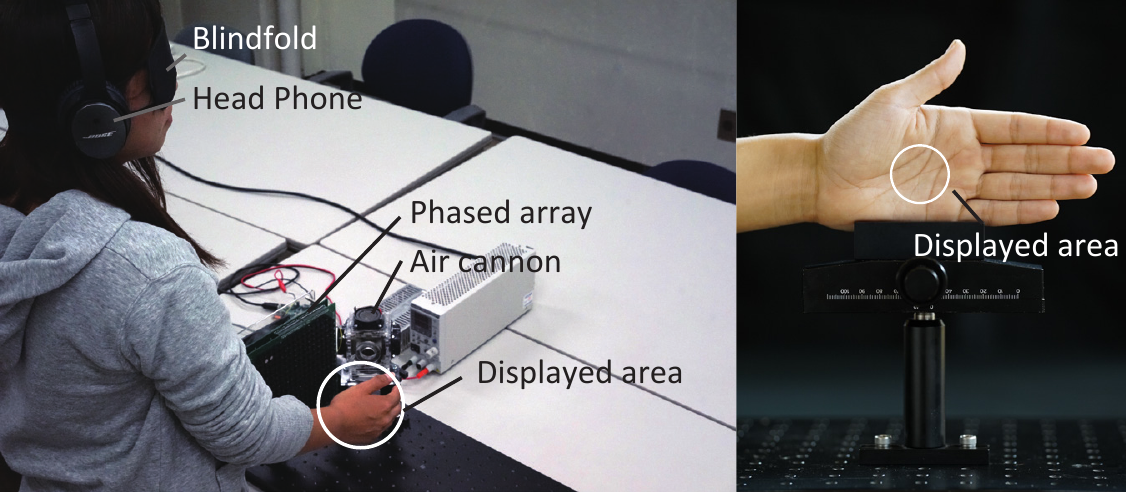}
  \caption{Overview of experimental setup}
\label{fig:expsetup}
 \end{figure}

The user study was conducted to evaluate the perceptual
threshold for acoustic radiation pressure induced by the focused
ultrasound and air the pressure vortex. Seven people participated in
the user study (19.6 years old on average with two females and five males).
The participants were isolated from visual information using 
 blindfolds, and auditory information was eliminated using
headphones with white noise (Figure \ref{fig:expsetup}, left).
Participants placed their hands on a table positioned 15 cm away from
the haptic device(Figure \ref{fig:expsetup} right).
The output force was set with reference to Figure \ref{fig:result_force}.

{\bf Focused ultrasound} : The focused ultrasound 
haptic stimulation was applied to the right palm of each participant. Moreover,  vibrotactile
stimulation modulated by 200-Hz and 50-Hz rectangular waves was applied.
The output force was set as one of six values (min:0.70 mN, max:10.9 mN)  near the
thresholds. The ultrasound output time was 200ms. Each force condition was applied once (i.e., one trial) and the
number of trials was 10 per participant. The order of trials was
randomized. In each trial,  the participants were asked whether they perceived stimuli on their
forefingers.

{\bf Aerodynamic vortex} : The aerodynamic vortex 
haptic stimulation was applied to the  right palm of each participant. 
The output force was set to one of six values (min:0.66 mN, max:13.7 mN) near the
thresholds. Each force condition was applied once (i.e., one trial) and the
number of trials was 10 per participant. The order of trials was
randomized. In each trial, the participants were asked whether they perceived the stimuli on their
forefingers.

{\bf Cross-field} : Three types of user studies were conducted on the
cross-field. First, focused ultrasound and an aerodynamic
vortex tactile stimulus were applied simultaneously. The output force of the air cannon was
set as 7.67 mN. The vibrotactile stimulation of ultrasound was set as 200 Hz
and 50 Hz. Second, the air vortex perceptual threshold in
space with a constant  ultrasound was evaluated. The output force of the ultrasound
was maintained at 9.7 mN. The output force of the air vortex was set
to one of six values (min:0.66 mN, max:13.7 mN). Third, the focused ultrasound perceptual threshold
in space with a constant
aerodynamic vortex was evaluated. The output force of the air
vortex was set as 7.67 mN and the air cannon was actuated to 20Hz. The
output force of ultrasound was set as one of six values (min:0.70 mN, max:10.9 mN) and the vibrotactile
stimulation was modulated to 200 Hz and 50 Hz.
  
 \begin{figure*}[tb]
\centering
 % \rule{17truecm}{8truecm}
  \includegraphics[width=0.8\linewidth]{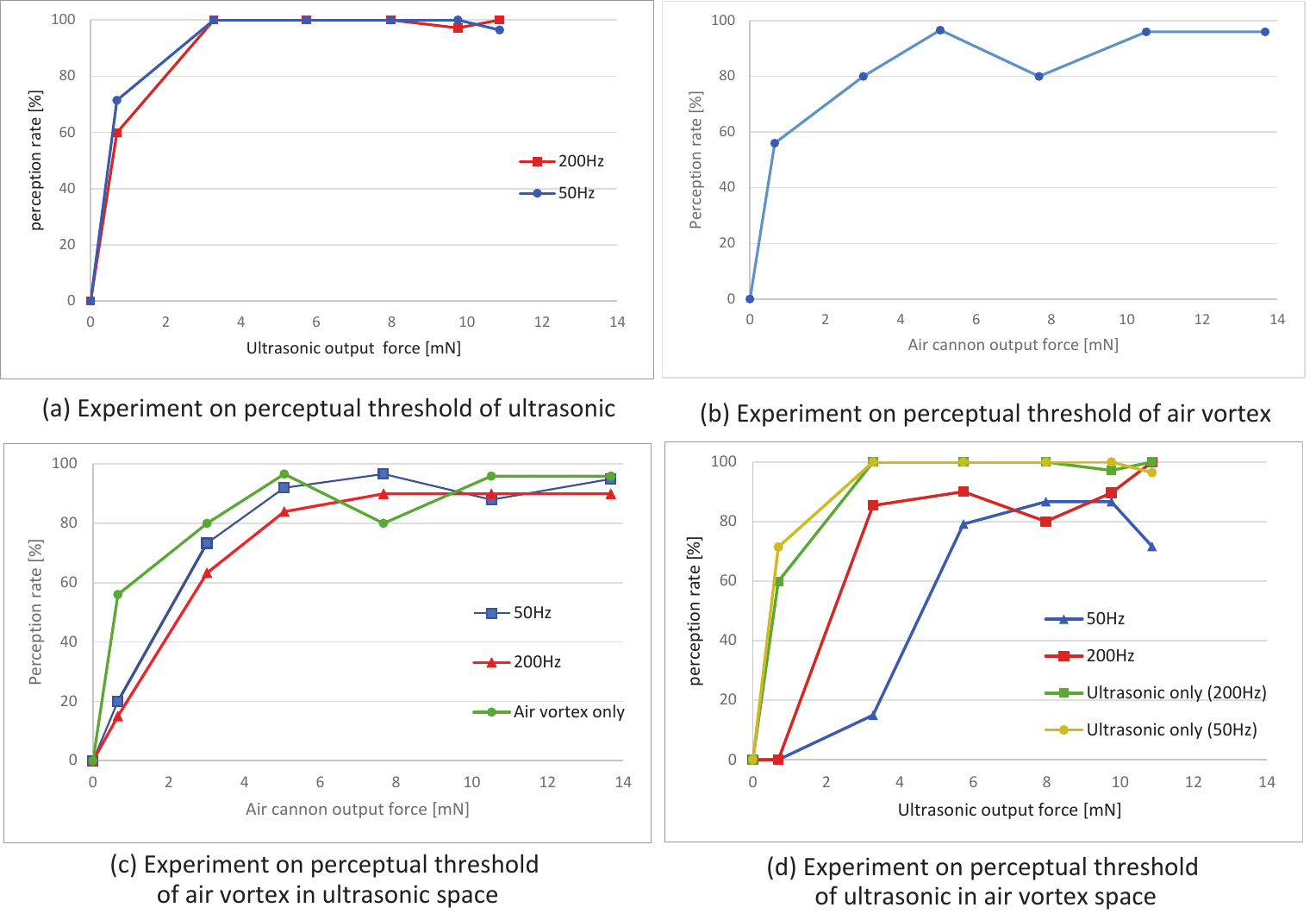}
  \caption{Results of perceptual threshold.}
\label{fig:result_threshold}
 \end{figure*}
\begin{table}[tb]
   \centering
\caption{Results of simultaneous tactile presentation with a focused ultrasound and  aerodynamic vortex}
\label{tab:result_threshold}
   \begin{tabular}{|c|c|} \hline
    The ultrasound  &The perceptual \\
    vibrotactile stimulation &  threshold of haptic \\
  {[Hz]} & [\%] \\ \hline
   50&95.2 \\ \hline
   200&100 \\ \hline
  \end{tabular}
  \end{table}
{\bf Results} : The results are presented in Figure \ref{fig:result_threshold} and Table \ref{tab:result_threshold}.
The perception rate on the vertical axis indicates the ease of tactile sensation. When the perception
threshold was 100\%, the examinee could perceive the tactile sensation.

In the case wherein only the ultrasound was applied, the perception threshold was nearly 100\% when the output exceeded 4 mN.
 In the case wherein only the air cannon was used, the perception threshold increased as the voltage increased. 
In addition, when the output force exceeded 11 mN, the participants could perceive feel the tactile sensation. 
When the air cannon was operated with an ultrasonic wave, sensing when the output force reached 2
 mN was difficult. 
However, when the output force was higher than 2 mN, the results were the same as that wherein only
 the air cannon was used. 
When ultrasonic waves were applied with an air cannon, the perception threshold was generally low
 compared with the case wherein only  ultrasonic waves was applied. 
In particular, when the modulation frequency was 50 Hz, the perception threshold was less than 20 \%. 
When presented simultaneously, participants could recognize a perception threshold of 95 \% or greater.

\section{Discussion and Conclusion}
From the experiments conducted on double-point thresholds, the double-point threshold was found to
increase the when air cannon and ultrasonic wave were combined. In addition, from the experiments on perceptual
thresholds, the  air cannon was easy to perceive in the ultrasonic presentation state; however,
the ultrasonic tactile stimulus was difficult to perceive in the air cannon presentation state. This is because the
magnitude of the force that the air cannon can apply is larger than the ultrasonic tactile stimuli.
It is necessary to appropriately adjust the magnitude of the applied force.
However, when the tactile stimulus was simultaneously applied, neither tactile sense could be perceived.
When using Sonovortex, it is effective to present the air cannon tactile sense in the ultrasonic
presentation state, or to simultaneously present the air cannon and the ultrasonic.

In this study, a method was developed that combines multiple tactile technologies.
This method generates a tactile sensation using an ultrasonic device and air cannon.
In addition, the ranges of possible resolutions and thresholds were discussed.
Cross-field helps eliminate the drawbacks of each field and 
provide a wider tactile presentation. An aerodynamic
vortex provides tactile sensations over large distances
and with considerable force levels. The focused ultrasound delivers distinguishable
high-resolution tactile sensations.

However, there are still several drawbacks. Both an air cannon and ultrasonic device generate environmental noise. 
In particular, given that the air cannon produces a loud sound, using it in a quiet space such as a
hospital or company office is difficult. 
However, this does not present a problem in noisy spaces such as shops and towns.
Moreover, to use Sonovortex, an air cannon and phased array must be deferred.
For wearing, Sonovortex is heavy and large. However, Sonovortex is expected to be incorporated into
the environment and used for digital signage and amusement.

\section*{Acknowledgement}
We would like to thank University of Tsukuba  for supporting this work. We are also
thankful to all the members of the Digital Nature Group at University of Tsukuba for their discussions and
feedback. 

\bibliographystyle{unsrt}
\bibliography{sono_v1}

\begin{thebibliography}{10}

\bibitem{Weigand1997}
A.~Weigand and M.~Gharib.
\newblock On the evolution of laminar vortex rings.
\newblock {\em Experiments in Fluids}, 22(6):447--457, 1997.

\bibitem{5406524}
T.~Hoshi, M.~Takahashi, T.~Iwamoto, and H.~Shinoda.
\newblock Noncontact tactile display based on radiation pressure of airborne
  ultrasound.
\newblock {\em IEEE Transactions on Haptics}, 3(3):155--165, July 2010.

\bibitem{Monnai:2014:HMH:2642918.2647407}
Yasuaki Monnai, Keisuke Hasegawa, Masahiro Fujiwara, Kazuma Yoshino, Seki
  Inoue, and Hiroyuki Shinoda.
\newblock Haptomime: Mid-air haptic interaction with a floating virtual screen.
\newblock In {\em Proceedings of the 27th Annual ACM Symposium on User
  Interface Software and Technology}, UIST '14, pages 663--667, New York, NY,
  USA, 2014. ACM.

\bibitem{Suzuki:2005:AJD:1042190.1042215}
Yuriko Suzuki and Minoru Kobayashi.
\newblock Air jet driven force feedback in virtual reality.
\newblock {\em IEEE Comput. Graph. Appl.}, 25(1):44--47, January 2005.

\bibitem{Carter:2013:UMM:2501988.2502018}
Tom Carter, Sue~Ann Seah, Benjamin Long, Bruce Drinkwater, and Sriram
  Subramanian.
\newblock Ultrahaptics: Multi-point mid-air haptic feedback for touch surfaces.
\newblock In {\em Proceedings of the 26th Annual ACM Symposium on User
  Interface Software and Technology}, UIST '13, pages 505--514, New York, NY,
  USA, 2013. ACM.

\bibitem{Long:2014:RVH:2661229.2661257}
Benjamin Long, Sue~Ann Seah, Tom Carter, and Sriram Subramanian.
\newblock Rendering volumetric haptic shapes in mid-air using ultrasound.
\newblock {\em ACM Trans. Graph.}, 33(6):181:1--181:10, November 2014.

\bibitem{Makino:2016:HMT:2858036.2858481}
Yasutoshi Makino, Yoshikazu Furuyama, Seki Inoue, and Hiroyuki Shinoda.
\newblock Haptoclone (haptic-optical clone) for mutual tele-environment by
  real-time 3d image transfer with midair force feedback.
\newblock In {\em Proceedings of the 2016 CHI Conference on Human Factors in
  Computing Systems}, CHI '16, pages 1980--1990, New York, NY, USA, 2016. ACM.

\bibitem{Suzuki:2002:DFF:506443.506608}
Yuriko Suzuki, Minoru Kobayashi, and Satoshi Ishibashi.
\newblock Design of force feedback utilizing air pressure toward untethered
  human interface.
\newblock In {\em CHI '02 Extended Abstracts on Human Factors in Computing
  Systems}, CHI EA '02, pages 808--809, New York, NY, USA, 2002. ACM.

\bibitem{Sodhi:2013:AIT:2461912.2462007}
Rajinder Sodhi, Ivan Poupyrev, Matthew Glisson, and Ali Israr.
\newblock Aireal: Interactive tactile experiences in free air.
\newblock {\em ACM Trans. Graph.}, 32(4):134:1--134:10, July 2013.

\bibitem{Gupta:2013:ANH:2493432.2493463}
Sidhant Gupta, Dan Morris, Shwetak~N. Patel, and Desney Tan.
\newblock Airwave: Non-contact haptic feedback using air vortex rings.
\newblock In {\em Proceedings of the 2013 ACM International Joint Conference on
  Pervasive and Ubiquitous Computing}, UbiComp '13, pages 419--428, New York,
  NY, USA, 2013. ACM.

\bibitem{Ueoka:2016:ICH:2851581.2892299}
Ryoko Ueoka, Mami Yamaguchi, and Yuka Sato.
\newblock Interactive cheek haptic display with air vortex rings for stress
  modification.
\newblock In {\em Proceedings of the 2016 CHI Conference Extended Abstracts on
  Human Factors in Computing Systems}, CHI EA '16, pages 1766--1771, New York,
  NY, USA, 2016. ACM.

\bibitem{Weiss:2011:FNH:2047196.2047277}
Malte Weiss, Chat Wacharamanotham, Simon Voelker, and Jan Borchers.
\newblock Fingerflux: Near-surface haptic feedback on tabletops.
\newblock In {\em Proceedings of the 24th Annual ACM Symposium on User
  Interface Software and Technology}, UIST '11, pages 615--620, New York, NY,
  USA, 2011. ACM.

\bibitem{Karunanayaka2013}
Kasun Karunanayaka, Sanath Siriwardana, Chamari Edirisinghe, Ryohei Nakatsu,
  and Ponnampalam Gopalakrishnakone.
\newblock {\em Magnetic Field Based Near Surface Haptic and Pointing
  Interface}, pages 601--609.
\newblock Springer Berlin Heidelberg, Berlin, Heidelberg, 2013.

\bibitem{7372383}
Q.~Zhang, H.~Dong, and A.~El Saddik.
\newblock Magnetic field control for haptic display: System design and
  simulation.
\newblock {\em IEEE Access}, 4:299--311, 2016.

\bibitem{6183773}
P.~Berkelman, M.~Miyasaka, and J.~Anderson.
\newblock Co-located 3d graphic and haptic display using electromagnetic
  levitation.
\newblock In {\em 2012 IEEE Haptics Symposium (HAPTICS)}, pages 77--81, March
  2012.

\bibitem{Saga2015}
Satoshi Saga.
\newblock {\em HeatHapt Thermal Radiation-Based Haptic Display}, pages
  105--107.
\newblock Springer Japan, Tokyo, 2015.

\bibitem{Ochiai:2016:FLF:2882845.2850414}
Yoichi Ochiai, Kota Kumagai, Takayuki Hoshi, Jun Rekimoto, Satoshi Hasegawa,
  and Yoshio Hayasaki.
\newblock Fairy lights in femtoseconds: Aerial and volumetric graphics rendered
  by focused femtosecond laser combined with computational holographic fields.
\newblock {\em ACM Trans. Graph.}, 35(2):17:1--17:14, February 2016.

\bibitem{Kyung:2008:WWG:1401615.1401657}
Ki-Uk Kyung and Jun-Young Lee.
\newblock wubi-pen: Windows graphical user interface interacting with haptic
  feedback stylus.
\newblock In {\em ACM SIGGRAPH 2008 New Tech Demos}, SIGGRAPH '08, pages
  42:1--42:4, New York, NY, USA, 2008. ACM.

\bibitem{5444646}
K.~Minamizawa, D.~Prattichizzo, and S.~Tachi.
\newblock Simplified design of haptic display by extending one-point
  kinesthetic feedback to multipoint tactile feedback.
\newblock In {\em 2010 IEEE Haptics Symposium}, pages 257--260, March 2010.

\bibitem{Lopes:2015:ISP:2807442.2807443}
Pedro Lopes, Alexandra Ion, and Patrick Baudisch.
\newblock Impacto: Simulating physical impact by combining tactile stimulation
  with electrical muscle stimulation.
\newblock In {\em Proceedings of the 28th Annual ACM Symposium on User
  Interface Software \&\#38; Technology}, UIST '15, pages 11--19, New York, NY,
  USA, 2015. ACM.

\bibitem{Ochiai:2016:CAH:2858036.2858489}
Yoichi Ochiai, Kota Kumagai, Takayuki Hoshi, Satoshi Hasegawa, and Yoshio
  Hayasaki.
\newblock Cross-field aerial haptics: Rendering haptic feedback in air with
  light and acoustic fields.
\newblock In {\em Proceedings of the 2016 CHI Conference on Human Factors in
  Computing Systems}, CHI '16, pages 3238--3247, New York, NY, USA, 2016. ACM.

\bibitem{7989930}
S.~{Hashizume}, K.~{Takazawa}, A.~{Koike}, and Y.~{Ochiai}.
\newblock Cross-field haptics: Multiple direction haptics combined with
  magnetic and electrostatic fields.
\newblock In {\em 2017 IEEE World Haptics Conference (WHC)}, pages 370--375,
  June 2017.

\bibitem{doi:10.1146/annurev.fl.24.010192.001315}
K~Shariff, , and A~Leonard.
\newblock Vortex rings.
\newblock {\em Annual Review of Fluid Mechanics}, 24(1):235--279, 1992.

\bibitem{:/content/aip/journal/pof1/31/12/10.1063/1.866920}
Ari Glezer.
\newblock The formation of vortex rings.
\newblock {\em Physics of Fluids}, 31(12), 1988.

\bibitem{dellon1987reliability}
A~Lee Dellon, Susan~E Mackinnon, and Page~McDonald Crosby.
\newblock Reliability of two-point discrimination measurements.
\newblock {\em The Journal of hand surgery}, 12(5):693--696, 1987.

\end{thebibliography}

\end{document}